# DIGITALISING THE WATER SECTOR: IMPLICATIONS FOR WATER SERVICE MANAGEMENT AND GOVERNANCE

Godfred Amankwaa, Global Development Institute, University of Manchester, godfred.amankwaa@manchester.ac.uk

Richard Heeks, Global Development Institute, University of Manchester, richard.heeks@manchester.ac.uk

Alison L Browne, Department of Geography, University of Manchester, alison.browne@manchester.ac.uk

**Abstract:** Digital technologies are becoming central to water governance and management, yet their impact and developmental implications are under-researched, particularly in the global South. This paper addresses this knowledge gap by examining the process of water service digitalisation and the resulting effects on service providers. Drawing on qualitative methods, we apply ideas on digitalisation, value, and power to investigate the implementation and impact of digital technologies in Ghana's state water utility company. We find digital water innovations to be recent, and delivering relatively limited impacts as yet, with value mainly accruing at the utility's operational rather than strategic level. The digital technologies present avenues for power shifts and struggles internally and externally as well as some changes in water management structures and responsibilities. We end with a brief discussion on the implications for water service governance and research.

**Keywords:** digital water, digitalisation, value, power, Ghana.

## 1. INTRODUCTION

The water sector increasingly needs innovation, and one important strand of that innovation has been digital water innovations (DWIs) (Daigger et al., 2019; Sarni et al., 2019). New technologies, ideas and approaches for digitising and "smartening" water systems are being embraced wholesale alongside the utilisation of data analytics in order to improve the sector (Hope et al., 2011; Cominola et al., 2015; Wade et al., 2020). Specific impacts claimed for DWIs include better service provision and a reconfiguration of the relationships between water users, providers and infrastructures (Guma et al., 2019; Hoolohan et al., 2021).

Although DWIs are in use in cities worldwide, arguments about their impact and developmental implications in the global South have to date been based on little research and evidence (Amankwaa et al., in review). As digital technologies are becoming a central feature of water governance in global South cities, important questions regarding how digitalisation and datafication transform, reproduce and reconfigure relations, power dynamics and knowledge systems within water service governance and management remain to be answered.

This paper aims to address this knowledge gap by focusing on the digitalisations associated with Ghana Water Company Limited (GWCL); Ghana's sole urban water service provider. Over recent years, GWCL has implemented a number of different digital water innovations and our analysis here provides new insights into the nature and limits to organisational value provided by these systems, and into change in organisational structures and power that have been associated with these new systems. By focusing on the impacts of digitilisation and datafication in a public water utility, this paper also adds to understanding of water and development, and to case material on the role of digital in development of the public sector in the global South.

The paper has five sections. In the following section, we briefly examine the literature on 'digitalisation, development and power' to understand the development impacts of digital in the water sectors. We discuss the progress of digitalisation in GWCL and the study methodology, and then present the study findings. The paper ends with the discussions and conclusions.

## 2. LINKING DIGITALISATION, DEVELOPMENT AND POWER

Digitalisation and its associated processes (e.g., datafication) are seen as integral to "smart city" visions across the world and have spread over the past decade to the global South through the aegis of local and global technology firms (Mayer-Schönberger & Cukier, 2013; Datta, 2015; Joss et al., 2015). In most global South countries, this agenda has focused on products such as data management systems for utilities and the application of digital technologies to urban development problems as ways of creating more efficient urban processes (Taylor & Richter, 2017; GSMA, 2020). Looking specifically at the water sector, we find digital technologies being implemented in both privatised and public water companies (Sarni et al., 2019).

In this paper, we focus on a public water utility (GWCL) to investigate the process of water services digitalisation and the resulting effects on service providers. To do this, we take inspiration from the concept of 'datafication and power' (Heeks et al., 2021) and logic of *epistemic determinism* (Cherlet, 2014) in order to explain how digitalisation and datafication relate to value, decision making and power.

In the literature on impact of digitalisation, two particular impact domains can be identified: *value and power*. *Value* deals particularly with the improvements that digital systems can make to organisational decisions. This is often discussed in terms of impacts across the organisational decision levels: operational, tactical and strategic (e.g., Turban et al. 2018). Value-related impacts of DWIs reported to date have typically been based on experiences in the global North or just on pilot or proof of concept experiences in the global South (Heymans et al., 2014; Ndaw, 2015; Monks et al., 2021). What evidence is available from the global South to date suggests limited value being derived from digital technologies for water service providers at either operational level such as improved water supply monitoring or more strategic value such as increase provider revenue (Hellstrom & Jacobson, 2014; Sarkar, 2019). Overall, however, there is so far little direct evidence about the impact of DWIs on decision-making and value in global South water provision (Amankwaa et al., under review).

The broader impact of digital systems includes how digital intersects with *power* and wider politics. Some studies have explained that the introduction of digital systems brings new actors and power dynamics in terms of "who counts", who has epistemic control and the implications of new structures and positionality of relations arising from digitalisation (Cherlet, 2014; Taylor & Broeders, 2015; Heeks et al., 2021). There is a very small amount of evidence of potential loss of power by groups or shifts in power and rights between groups in developing countries through digital systems in public utilities (Heeks et al., 2021). Others have argued that digital systems may empower some actors such as large corporations or the state at the expense of others (Hilbert, 2016). In the water sector however, evidence-based analysis of broader impacts and power relations of digitalisation is much more limited, especially among water service providers in the global South (Taylor & Ritchie, 2017). There is little evidence and exploration of how the context of interests and power shapes DWI implementation and the complementary changes in capabilities, incentives and management processes it might bring within the water service governance and management (Amankwaa et al., under review).

In sum, we know from past literature that digital water innovations have the potential to impact both value and power in global South water systems, but we have as yet too little evidence on this. We now move to explain how this current study sought to illuminate these issues.

# 3. DATA AND METHODS

## 3.1. Research setting

This study focused on Ghana based on the current spate of digital systems roll-out within the water sector as part of both national and water service provider goals (World Bank, 2019; Amankwaa et al., 2020). To address the research aims, we employed a case study approach to enable in-depth research of digital technologies and their impact (Yin, 2014). The chosen case was that of the Ghana Water Company Limited. GWCL is the sole government-owned water utility company in the country, and it is responsible for the production, transmission and distribution of water in urban areas in Ghana. The Company manages 90 water systems serving about 11 million people nation-wide with headquarters in Accra alongside 15 regional and 90 district offices (GWCL, 2019).

Alongside its critical role in Ghana's water sector, GWCL was also selected because it has been following digital road map, implementing digital water innovations in its operations from source to consumer (GWCL, 2019; Waldron et al., 2019; Amankwaa et al., 2020). As summarised in Table 1, these relate to three main domains. Distribution and delivery applications have mapped and modelled the Company's water infrastructure and monitor water flow and pressure. User-related systems monitor water use levels and digitise billing, payment and other interactions with consumers. Internally, an enterprise resource planning (ERP) system is in operation across procurement, HR, accounting and other departments.

| Stages of the water value chain | Activities | Key digital water innovations implemented |
|---|---|---|
| Distribution and delivery | Water distribution and monitoring | - Geographic information system<br>- Hydraulic modelling<br>- Sensors (SCADA (supervisory control and data acquisition), loggers, etc.)<br>- Telemetry system |
| End-users | - Water consumption<br>- Consumption monitoring and revenue collection<br>- Service feedback | - Smart (ultrasonic) meters with 'drive by' radio technology<br>- Electronic billing and payment system<br>- Online customer feedback services (e.g., customer mobile apps and portal)<br>- Customer engagement channels including call systems, Telegram and WhatsApp |
| Internal | Procurement (Materials), HR, accounting, etc | - ERP system |

**Table 1. Digital water innovations implemented by GWCL.**

Implementation of these digital systems has been slowed by both human and technical factors. However, these have not prevented the implementation. Digital water innovations may still be at a relatively early stage, but they are sufficiently embedded to make a study of their impacts practicable: another rationale for the selection of GWCL as a case study.

### 3.2. Methodology

Within the case study strategy, the research reported here adopted qualitative methods comprising key informant interviews and policy document analysis. Initial interviewee sampling was purposive and snowballed, beginning with pre-existing contacts in GWCL and from them identifying those within GWCL involved with decision-making processes at different levels in the organisation. Contacts with external experts were snowballed to add others based on recommendations. In all, the first author conducted 16 semi-structured interviews from November 2020 to February 2021, with 12 GWCL professionals (senior officers and mid- and operational-level staff), one representative of an international water organisation who had previously worked for a water-related civil society organisation in Ghana, two GWCL external partners, and one consultant of USAID who was working on public utility digital water innovations in Ghana. These people were selected in order to triangulate evidence sources and because of their explicit knowledge on the subject of digital innovations in the urban water sector.

The interviews were designed around issues identified from the literature: the type and nature of technology implemented, drivers of implementation, issues around value created, and broader impacts associated with digitalisation and datafication. Due to Covid-19 restrictions, interviews were conducted online using Microsoft Teams and Zoom or via phone calls. All interviews were conducted in English and they were of 30–50 minutes' duration. Interviews were audio-recorded, or the researcher took handwritten notes where respondents were not comfortable with recording. Company reports were sourced from GWCL and broader policy documents on digital and water in Ghana were downloaded from government, international, civil society and other organisations.

Data from the different sources were transcribed as text, coded and thematically analysed by comparing the different responses in order to identify common trends, themes, similarities and contrasts. Coding and analysis were done using key dimensions highlighted in the literature – particularly value and power – as a starting point. Once new themes emerged, iterations were made and coded as part of a revised coding frame.

## 4. FINDINGS

### 4.1 Organisational value and impact of digitalisation

The organisational value of digital water innovations can be analysed in relation to GWCL's different areas of activity.

*Faults and leakages* of the main water distribution system used to be detected via customer reportage and perhaps field officers. According to a GWCL Engineer, this is still the case for smaller distribution lines and the end-pipes linked to domestic and other user properties. However, main (transmission) boundary pipeline networks of the operational areas in Accra and the larger distribution lines of the network now use a combination of SCADA sensors and telemetry systems to automatically report falls in water pressure or faults, with instant alarm notifications sounding in GWCL's central control room. When this happens, an instrumentation engineer in the control room can identify the likely location of the problem. In some cases, it can now be resolved remotely or, if a field officer needs to be despatched, they can be directed to the likely location so that fault location and resolution is now more efficient (GWCL Engineer 4). All of this is facilitated by mapping of the network which, as a GWCL Technician explained, has given location IDs to all main elements of the water network including customer locations. In sum, then, digitalisation has been associated with better and faster monitoring, fault location and resolution.

These deployments also offer real-time data for the *day-to-day operation* of water systems. A GWCL Engineer indicated that, "…the SCADA systems and boundaries gives us clear picture on how much water is going in and out of the three regions: Tema, Accra West and Accra East. This offers a basis on where to ration water to or not on daily basis". Likewise, a GWCL Meter Technician explained that, "data from Telemetry and SCADA has helped them allocate water and ration water to regions and know how much water is distributed to particular regions". The digital technologies are therefore allowing a better understanding and intervention in terms of routing and rationing of water in the Company's different regions of operation.

General *billing* was reported to be more accurate and faster than previously. Previously, analogue meters might not be present at all or if present might not actually be read – so-called "armchair meter reading" – with the entered number being some mix of guesswork, intuition and experience of the meter readers (GWCL Engineer 1). For large-scale users then usage is recorded automatically: "using bulk meters with sensors has informed us the right amount of water a particular industrial customer uses to enable accurate billing" (GWCL Meter Technician 1). For smaller-scale including domestic consumers, meters are read using handheld devices with an eReader app. This includes the option to take a picture of the meter which can be referred to later in case of any query or challenge to the reading, and it also allows remote co-monitoring of readings by supervisors and middle managers. Although the direct causal link to digital technology is hard to establish it is noted that a marginal increase in billing performance has occurred in recent years with, for example, an increase of 4.2% in 2018 compared to 2017 (GWCL, 2019).

*Payment* has also been digitalised with the option for customers of mobile and other digital payment channels, linked to the ability of customers to receive their bills digitally via SMS or email (GWCL Regional Officer 1; Amankwaa et al., 2020). Internal cost-benefit analysis indicates that using digital payment channels rather than cash payments saves the company about 10% of the administration cost of collecting and processing bills (DWI Consultant). Linking digital technology to wider impacts is difficult. However, GWCL saw a 14.1% increase in revenue for the year 2018 (GWCL, 2019) and GWCL officials largely attribute this to the introduction of digital billing and payment systems (GWCL Engineer 1).

Finally, *customer engagement* systems have been digitised with a call centre management system and use of messaging platforms like Telegram and WhatsApp. One indicator of value was that the number of customer complaints at the district level was said to have decreased because of these systems, and rectification time is reported to have also decreased significantly (GWCL District Officer 1; GWCL, 2019).

Looking at the overall pattern, the first finding would be that digital water innovations do appear to be adding value to the operations of this public water utility. But "operations" is the key term. If we look in terms of operational, tactical and strategic levels then almost all of the value to date is being derived in terms of operational decisions and processes. While these may aggregate up to impact organisational performance indicators – perhaps, for example, in terms of operational revenue – there is limited evidence yet of impact on higher-level decisions and processes. For example, we did not yet find an impact on the typical strategic decisions required of water utilities such as risk analysis, infrastructure expansion and development, or corporate reform (MacGillivray et al., 2006; Mugabi et al., 2007; Mutikanga et al., 2011). This concentration of digital value at the lower levels of the organisation may reflect the relative recency of digitalisation. Perhaps linked to this also seems to reflect a lack of understanding in the organisation of the strategic value of the data being generated as a result. As one example, we were unable to identify anyone within GWCL whose role was to process and analyse the new digital data within the organisation and present it to middle and senior managers for their use.

**4.2 Powershifts, digital water politics and water governance**

The first component of broader impact of digitalisation found was change in some *organisational structures and responsibilities*. We found that there has been a centralisation of power and focus around digital in the organisation. A Technology and Innovation Department (T & I) had been expanded and given sole responsibility for overseeing all issues relating to digital transformation within GWCL: not just implementation but also strategic decision-making (GWCL Engineer 2). This integrates a whole series of previous distributed and separate responsibilities, covering three technology-related bodies – the Geographic Information Systems Department, the ICT Department, and the Metering, Instrumentation and Non-Revenue Water Reduction Department – plus the Research unit (GWCL, 2019).

Given the growing role of digital within GWCL, T & I has increasing outreach and connections with the rest of the organisation. Because of the introduction of smart meters, for example, Engineers and officials in the T & I department link out to meter readers and technicians; providing support and running training programmes. Because of the sensor and telemetry systems, they support and train field officers. Because of the growing role of digital in operations, they support and train regional and district managers. These were tasks not previously undertaken and/or not previously centralised in one department being the responsibility of commercial and operations departments in the organisation (GWCL Engineer 1). Where the scope and reach of earlier incarnations would have been restricted largely to corporate headquarters, "T & I managers and officers have now become central points in dealing with most technical field complaints associated with digital technologies such as meter reading" (GWCL Engineer 1). An institutional equivalent of the central control room with its panopticon-style overview of the whole water distribution network, the T & I Department now has a digital overview of the whole organisation, linked through the threads of its digital systems to every part of GWCL and its operations.

The other impact-related finding concerns the *powershifts* and struggles among actors. Digital water innovations have been associated with some shifts in the locus of *power* within and outside the utility. First, there has been an upward shift in power to management from operational staff (mainly meter readers and some technicians). Previously, these staff lay largely beyond the direct gaze of management, but this is no longer the case. For instance, middle and even senior managers can now monitor how lower-level staff operate and get data on their performance. Even the Managing Director of the company can now monitor the operational performance of all of meter readers (GWCL Meter Reader 2). A meter reader recounted: "I got my appraisal delayed because my performance information on the system indicated I hadn't achieved the 100% targets for the past three months. Data about work and our information is everywhere even to the Chief Managers, so there's no room for shoddy works". Middle- and senior-level managers have thus gained greater power through their access to and capacity to use digital data as a managerial and epistemic resource within the organisation. Via digital technology, they have been able to cut through organisational layers that previously interceded between top and bottom of the organisation.

Second, there has been a shift of some form of power to private sector actors though this has continuously been contested as part of a historical pattern. This began shortly after the turn of the century when Indian company Aquamet was issued a contract in 2004 to supply, install and collect revenue from prepaid water meters (Shang-Quartey, 2017). Though the digital component of this project was limited, technical faults along with the clash between profit and public welfare logics led to the contract being abandoned. The connection between digital and privatisation continued, however, with a management contract being issued to Aqua Vitens Rand Ltd in 2006. During the five-year period of the contract, the foundations were laid for many of the digitalisations described above including metering and mapping and customer engagement (Abubakari et al., 2013). However, due to human and technical implementation issues, these did not deliver the desired impacts and – driven much more by wider failure to improve water services and conflict between public and private sector culture and objectives – the contract was not renewed when it ended in 2011. These experiences stymied a further attempt in 2014 and 2015 to roll out prepaid meters. Despite this being a GWCL initiative, huge opposition by civil society organisations – shaped by past experiences of private operators and feeling that these meters "contributed to attempts in privatising 'public' water in the country" (Former Civil Society Organisation Coordinator) – led to these initiatives being abandoned.

More recently, implementation of the electronic billing project gave some form of *de facto* control of the system to a private operator; SOFTtribe. SOFTtribe is a Ghanaian software developing company that was contracted to develop, manage, operate and provide data integration services for GWCL's e-billing and e-payment systems (DWI Consultant). SOFTtribe's control over key aspects of the system after about two years led to disagreements on the operationalisation, management and ownership of the system and its related data. This ultimately led to termination of the contract between GWCL and SOFTtribe. This was therefore just the most-recent example of concerns about the way in which private sector deployment of digital systems has led to a loss of power and control from public to private sectors. This shift has derived from the power of the rights, processes and resources including data and knowledge that are bound up into digital systems; those powers increasing as digitalisation spreads within the organisation.

A third potential power shift could be between the public utility and its customers. When asked how customers are represented in the digital water value chain, GWCL engineers and meter readers held the view that technologies such as smart meters offer customers elements of greater operational transparency and service benefits. For example, smart meters enable customers to just pay for "what they consume" and afford them the data resources necessary for them to monitor and challenge water bills (GWCL Meter Reader 3). But conversely, customers and their actions become more transparent to GWCL, with the Company now knowing locations, accurate usage levels and more about their customers.

## 5. DISCUSSION AND CONCLUSIONS

In this study, our overall interest was to understand the impact of water service digitalisation on water service providers in the global South. We focused on a public utility to understand how digitalisation and datafication impact the water service governance and management. Two core findings emerged which forms the basis for discussion. First, the impact of digital water innovations is not yet transformative: they have had limited impact in more strategic terms but are already delivering value at the operational level. Second, digital technologies present new avenues for power shifts and struggles between and within organisations as well as changes in organisational structures and responsibilities in water governance.

A few years into implementation of a programme of digital transformation, we found that value from digital systems was emerging slowly within GWCL and may be making some contribution to a few organisational goals. While some literature assumes that digital water systems will be applied to strategic and outward-facing purposes (Antzoulatos et al., 2020), in GWCL they were rather focused on operational level activities and decisions such as using technologies for accurate meter reading, pipeline monitoring, and producing real-time data for the day-to-day operation of water systems. We found some potential evidence of financial value, with the internal estimate of 10% cost savings in relation to the billing and payment systems. It is beyond the scope of this study to undertake cost-benefit analysis but there is potential that digital systems may make a positive contribution to the financial bottom line, particularly as growing numbers of customers adopt digital payment. Studies have reported similar benefits in the global North (Beal & Flynn, 2015; March et al., 2017). However, literature on actual benefits in global South water systems is very limited, and our findings therefore provide a first set of systematic, real-world evidence on the organisational value of digital water innovations.

Aside from these operational-level impacts, digital technologies have not yet been transformative as there is little sign of value at the strategic level and in terms of organisation-wide performance indicators. One indicator of this was the lack of strategic value extraction from the great deal of data now being generated by GWCL's new digital systems. Overall, therefore, this study questions the narrative of "digital water transformation" (Alabi et al., 2019; Hoolohan et al., 2021) and instead highlights the much more incremental value being delivered by digital water systems.

Within the digital water literature, there are arguments that digitalisation is associated with a reconfiguration of the relationships and power relations between stakeholders in the water sector (Guma et al., 2019; Hoolohan et al., 2021). Evidence from our study emerged to not only confirm the existence of such reconfigurations but also to provide a first clear mapping of them in a global South context. For instance, it emerged that, internally, digital systems have brought about changes in organisational structures and responsibilities in water governance. Contrary to the findings of earlier literature (Owen, 2018; Sarni et al., 2019), we found that – rather than automating human labour – the focus of those interviewed was the way in which digital systems increased the workload of some utility workforce especially those at the forefront of digital operations. As digitalisation introduces new equipment and new systems to the water sector, the emphasis here – as predicted by Wallis & Johnson (2020) – has been a need to employ new expertise and talent.

The rationality and automation associated with digital systems might suggest they could eliminate or reduce power struggles. The findings suggest the opposite, however: that these technologies have been associated with and even triggered power shifts and struggles within GWCL and with external partners such as Aquamet and SOFTtribe. Internally, these relate to the power provided by digital data and the transparency of actors and actions this provides. Externally, that some power leads to conflict over data systems ownership and control. The findings are in line with those of earlier literature which links digital water systems to shifts in power from public to private sector (Taylor & Richter, 2017), and to similar findings in other public utilities (Heeks et al., 2021).

Alongside the power dynamics and shifts we have seen associated with digital water systems, there are wider questions about who is empowered or excluded by these powershifts and struggles. In our study, we found that central management and certain central departments (e.g., T & I) within the company seem to have epistemic control over different aspects of the digital water system. Internally, then, digital systems have provided central management and supervisors with direct monitoring and epistemic control over the activities of some field and other lower-level staff. The latter have thus been disempowered in relative terms. While it did not explicitly emerge from interviews, the centralisation and relative empowerment of the T & I Department is likely to lead to tensions with other departments about who leads new digital initiatives and who controls digital systems and their related powers.

This discussion reinforces previous analyses that suggest digital infrastructures and technologies perform (power-related) political work with important consequences for water governance (von Schnitzler, 2016; Guma et al., 2019; Hoolohan et al., 2021). Insights into the power relations and digital water politics are important for the design, implementation, and governance of digital systems and water services. Therefore, the analysis here will be vital for understanding how digitalisation and datafication transform, reproduce and reconfigure relations, power dynamics and knowledge systems within the water sector.

## 5.1 Conclusions

This paper has provided real-world case evidence and empirical insights into the impacts and implications associated with water service digitalisation in the global South. It has also responded to recent calls by scholars like Hoolohan et al. (2021) and Amankwaa et al. (in review) on the need for systematic examination of the impacts of digital water innovations within a wider socio-technical and social political lens. Both water researchers and practitioners need to recognise both the value and political impacts of digital water innovations. For researchers, more work on these issues is required including analysis of digital systems over time; for example, to see if they start to have more strategic and transformative impacts within water service providers, and externally to understand more fully the way that digital impacts power balances and relations with external stakeholders including customers. Water service providers need to understand the "value gap" between the impact digital systems could have and what they currently have; for example, seeking more ways to extract strategic value from the datafication these systems enable. They need to grasp the politics of digital; seeing that these systems cannot simply be understood in technocratic terms, and particularly understanding how digital may change the relationship with their customers. There is the need, for example, more inclusive models of water management that could be applied to the growing diffusion of digital water systems; models that consider the everyday realities of end-users in global South cities.

# REFERENCES

Abubakari, M., Buabeng, T., & Ahenkan, A. (2013). Implementing public private partnerships in Africa: the case of urban water service delivery in Ghana. Journal of Public Administration and Governance, 3(1), 41-56.

Alabi, M., Telukdarie, A., & Jansen, N. V. R. (2019). Industry 4.0 and Water Industry. In Proceedings of the International Annual Conference of the American Society for Engineering Management. (pp. 1-11). American Society for Engineering Management (ASEM).

Amankwaa, G., Asaaga, F. A., Fischer, C., & Awotwe, P. (2020). Diffusion of electronic water payment innovations in urban Ghana. Evidence from Tema Metropolis. Water, 12(4), 1011.

Amankwaa, G., Heeks, A., & Browne, A.L. (n'd). Digital Innovations and Water Services in Global South Cities: A Systematic Literature Review

Antzoulatos, G., Mourtzios, C., Stournara, P., Kouloglou, I. O., Papadimitriou, N., Spyrou, D., ... & Kompatsiaris, I. (2020). Making urban water smart: the SMART-WATER solution. Water Science and Technology, 82(12), 2691-2710.

Beal, C. D., & Flynn, J. (2015). Toward the digital water age: Survey and case studies of Australian water utility smart-metering programs. Utilities Policy, 32, 29-37.

Cherlet, J. (2014). Epistemic and technological determinism in development aid. Science, Technology, & Human Values, 39(6), 773-794.

Cominola, A., Giuliani, M., Piga, D., Castelletti, A., & Rizzoli, A. E. (2015). Benefits and challenges of using smart meters for advancing residential water demand modeling and management: A review. Environmental Modelling & Software, 72, 198-214.

Datta, A. (2015). New urban utopias of postcolonial India: 'Entrepreneurial urbanization' in Dholera smart city, Gujarat. Dialogues in Human Geography, 5(1), 3-22.

Daigger, G. T., Voutchkov, N., Lall, U., & Sarni, W. (2019). The future of water. A collection of essays on "disruptive" technologies that may transform the water sector in the next 10 years. Discussion paper No. IDB-DP-657. Washington, DC: Inter-American Development Bank.

GSMA (2020). Digital solutions for the Urban Poor. London, UK: GSMA

Guma, P. K., Monstadt, J., & Schramm, S. (2019). Hybrid constellations of water access in the digital age: the case of Jisomee Mita in Soweto-Kayole, Nairobi. Water Alternatives, 12(2), 725-743.

GWCL. (2018). GWCL e-Billing Project: Progress Report; Number 21.12.07.18; Ghana Water Company Limited: Accra, Ghana, 2018

GWCL. (2019). 2018 Annual Performance Report. Ghana Water Company Limited: Accra, Ghana.

Heeks, R., Rakesh, V., Sengupta, R., Chattapadhyay, S., & Foster, C. (2021). Datafication, value and power in developing countries: Big data in two Indian public service organizations. Development Policy Review, 39(1), 82-102.

Hellström, J., & Jacobson, M. (2014). 'You Can't Cheat the Community Anymore'–Using Mobiles to Improve Water Governance. In Proceedings of 4th International Conference on M4D Mobile Communication for Development (pp. 48-59). Karlstad University Studies, Karlstad.

Heymans, C., Eales, K., & Franceys, R. (2014). The Limits and Possibilities of Prepaid Water in Urban Africa. Washington, DC: World Bank Group.

Hoolohan, C., Amankwaa, G., Browne, A. L., Clear, A., Holstead, K., Machen, R., ... & Ward, S. (2021). Resocializing digital water transformations: Outlining social science perspectives on the digital water journey. Wiley Interdisciplinary Reviews: Water, e1512.

Hope, R., Foster, T., Money, A.; Rouse, M., Money, N., & Thomas, M. (2011). Smart Water Systems; Project Report to UK DFID; Oxford University: Oxford, UK.

Joss, S., Sengers, F., Schraven, D., Caprotti, F., & Dayot, Y. (2019). The smart city as global discourse: Storylines and critical junctures across 27 cities. Journal of urban technology, 26(1), 3-34.

MacGillivray, B. H., Hamilton, P. D., Strutt, J. E., & Pollard, S. J. (2006). Risk analysis strategies in the water utility sector: An inventory of applications for better and more credible decision making. Critical Reviews in Environmental Science and Technology, 36(2), 85-139.

Monks, I. R., Stewart, R. A., Sahin, O., Keller, R. J., & Prevos, P. (2021). Towards understanding the anticipated customer benefits of digital water metering. Urban Water Journal, 1-14.

Mugabi, J., Kayaga, S., & Njiru, C. (2007). Strategic planning for water utilities in developing countries. Utilities policy, 15(1), 1-8.

Mutikanga, H. E., Sharma, S. K., & Vairavamoorthy, K. (2011). Multi-criteria decision analysis: a strategic planning tool for water loss management. Water resources management, 25(14), 3947-3969.

Taylor, L., & Broeders, D. (2015). In the name of Development: Power, profit and the datafication of the global South. Geoforum, 64, 229-237.

Taylor, L., & Richter, C. (2017). The power of smart solutions: Knowledge, citizenship, and the datafication of Bangalore's water supply. Television & New Media, 18(8), 721-733.

Turban, E., Pollard, C., & Wood, G. (2018). Information Technology for Management: On-demand Strategies for Performance, Growth and Sustainability. John Wiley & Sons.

Shang-Quartey. L. (2017). The fall of prepaid water meters in Ghana. An account of civil society organizations' campaign for human right to water. Water Citizens Network of Ghana

Sarkar, A. (2019). The role of new 'Smart technology' to provide water to the urban poor: a case study of water ATMs in Delhi, India. Energy, Ecology and Environment, 4(4), 166-174.

Sarni, W., White, C., Webb, R., Cross, K., & Glotzbach, R. (2019). Digital water: Industry leaders chart the transformation journey. London, UK: International Water Association White Paper.

Mayer-Schönberger, V., & Cukier, K. (2013). Big data: A revolution that will transform how we live, work, and think. Houghton Mifflin Harcourt.

Von Schnitzler, A. (2016). Democracy's infrastructure: Techno-politics and protest after apartheid. Princeton University Press.

Wallis, T. & Johnson, C. (2020). Implementing the NIS Directive, driving cybersecurity improvements for essential services. In International conference on cyber situational awareness, data analytics and assessment (CyberSA), Dublin, Ireland. 1–10.

Wade, M. J., Steyer, J. P., & Garcia, M. V. R. (2020). Making water smart. Water Science and Technology, 82(12), v–vii.

World Bank. (2019). Ghana Digital Economy Diagnostic: Stocktaking Report. The World Bank Group, Washington DC, USA.

Yin, R. K. (2014). Case study research: Design and methods (5th ed.). Los Angeles, CA: Sage